\documentclass[12pt]{article}

\usepackage{scicite}

\usepackage{times}
\usepackage{chemformula} 
\usepackage[T1]{fontenc} 
\usepackage{mciteplus}
\usepackage{float}
\usepackage[utf8]{inputenc}

\newenvironment{sciabstract}{%
\begin{quote} \bf}
{\end{quote}}

\title{Synthesis of an aqueous, air-stable, superconducting 1T'-\ch{WS2} monolayer-ink} 

\author
{Xiaoyu Song,$^{1}$ Ratnadwip Singha,$^{1}$ Guangming Cheng,$^{2}$ Yao-Wen Yeh,$^{3}$ \\Franziska Kamm,$^{4}$ Jason F. Khoury,$^{1}$ Florian Pielnhofer,$^{4}$ \\Philip E. Batson,$^{3}$ Nan Yao,$^{2}$ Leslie M. Schoop$^{1\ast}$\\
\\
\normalsize{$^{1}$Department of Chemistry, Princeton University, Princeton, NJ 08544, USA}\\
\normalsize{$^{2}$Princeton Institute for Science and Technology of Materials, Princeton, NJ 08544, USA}\\
\normalsize{$^{3}$Department of Physics \& Astronomy, Rutgers University, Piscataway, NJ 08854, USA}\\
\normalsize{$^{4}$Institute of Inorganic Chemistry, University of Regensburg, D-93040 Regensburg, Germany}\\
\normalsize{$^\ast$To whom correspondence should be addressed; E-mail:  lschoop@princeton.edu.}
}

\date{}

\begin{document} 


\baselineskip24pt


\maketitle

\begin{sciabstract}
Liquid-phase chemical exfoliation is ideal to achieve industry scale production of two-dimensional (2D) materials for a wide range of application such as printable electronics, catalysis and energy storage. However, many impactful 2D materials with potentials in quantum technologies can only be studied in lab settings due to their air-sensitivity, and loss of physical performance after chemical processing. Here, we report a simple chemical exfoliation method to create a stable, aqueous, surfactant-free, superconducting ink containing phase-pure 1T'-\ch{WS2} monolayers that are isotructural to the air-sensitive topological insulator 1T'-\ch{WTe2}. We demonstrate that thin films can be cast on both hard and flexible substrates. The printed film is metallic at room temperature and superconducting below 7.3 K, shows strong anisotropic unconventional superconducting behavior with an in-plane and out-of-plane upper critical magnetic field of 30.1 T and 5.3 T, has a critical current of 44 mA, and is stable at ambient conditions for at least 30 days. Our results show that chemical processing can provide an engineering solution, which makes non-trivial 2D materials that used to be only studied in laboratories commercially accessible.  

\end{sciabstract}

\section*{Introduction}

\begin{figure}[H]
\centering
  \includegraphics[width=1\textwidth]{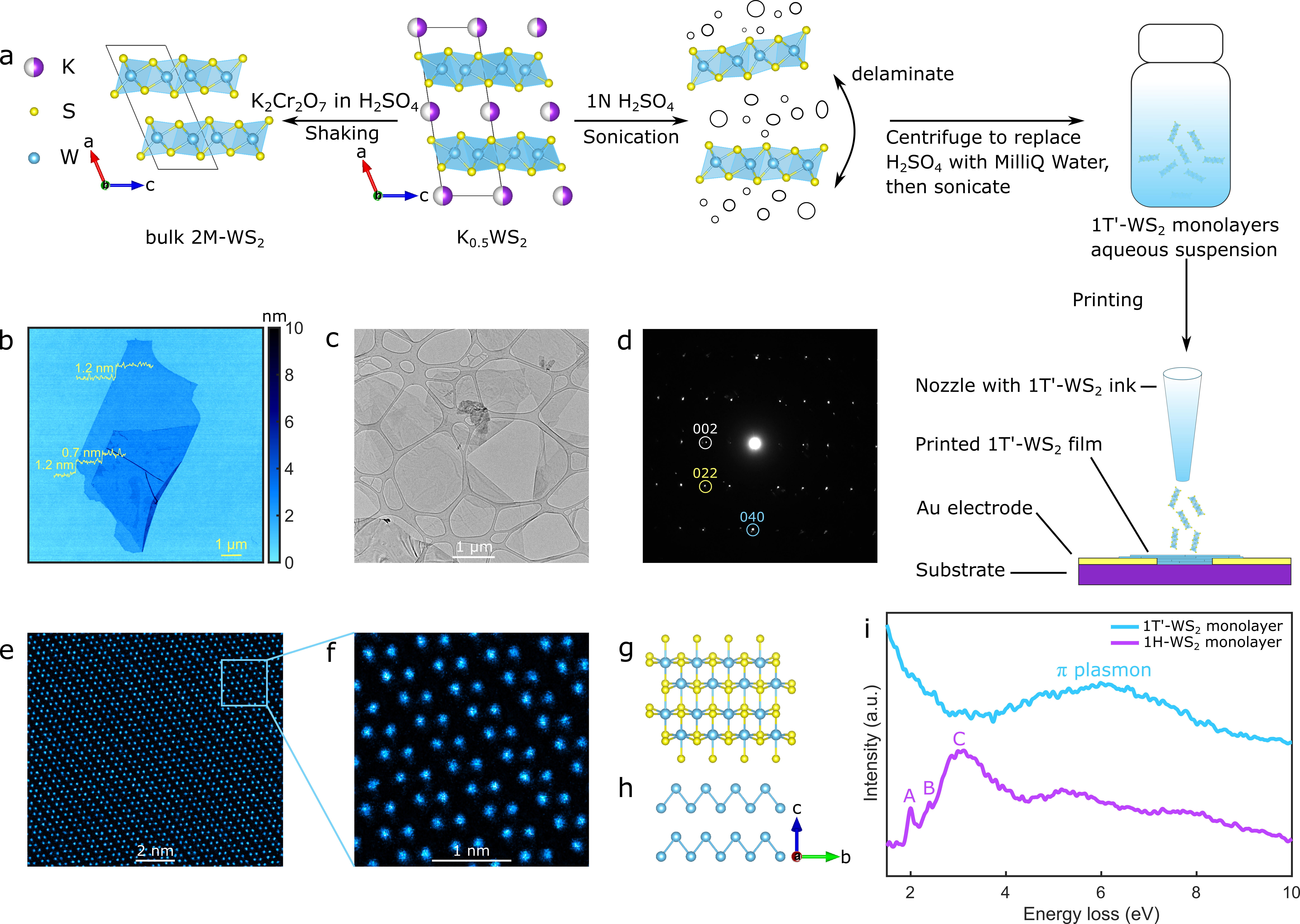}
  \caption{(a) A schematic explaining the synthesis of bulk 2M-\ch{WS2} made of layered face-sharing distorted \ch{WS6} octahedra, as well as an aqueous  1T'-\ch{WS2} nanosheet-ink from \ch{K_{0.5}WS2}. (b) An atomic force microscopy (AFM) image of a monolayer 1T'-\ch{WS2}. (c) A transmission electron microscopy (TEM) image of 1T'-\ch{WS2} monolayers. (d) Selected area electron diffraction (SAED) of a free-standing 1T'-\ch{WS2} monolayer. (e) An atomic resolution scanning transmission electron microscopy (STEM) image of a monolayer 1T'-\ch{WS2}, where W atoms are highlighted with blue color. (f) A zoom-in atomic-resolution image of a 1T'-\ch{WS2} monolayer. (g) In-plane crystal structure of monolayer 1T'-\ch{WS2}. The zigzag chains of W atoms are shown in (h). (i) Electron energy loss spectroscopy (EELS) of a 1T'-\ch{WS2} monolayer in comparison with a 1H-\ch{WS2} monolayer.}
  \label{figure1}
  \centering
\end{figure}

Two-dimension (2D) materials offer opportunities towards the discovery of new physics as well as novel type of applications, such as in flexible or wearable electronics\cite{torrisi2014electrifying,lin2018solution,eda2008large,shi2017highly,kelly2021electrical,li2021printable}. One exciting material is monolayer \ch{WTe2}, which is a 2D topological insulator (TI) with an excitonic insulator ground state that becomes superconducting upon gating\cite{wu2018observation,fatemi2018electrically,sajadi2018gate}. Monolayer 1T'-\ch{WS2} had been predicted to be a 2D TI with a larger gap\cite{qian2014quantum}, but its synthesis is challenging, as \ch{WS2} prefers to adopt the more stable semiconducting 2H phase (Fig. S1 a,b), which can be directly exfoliated into semiconducting 1H-\ch{WS2} monolayers (Fig. S1 c,d) by direct sonication in various organic solvents\cite{coleman2011two}. Recently, a new phase of \ch{WS2} was reported which brought the synthesis of 1T'-\ch{WS2} monolayers one step closer: By oxidizing \ch{K_{0.7}WS2} with either \ch{K2Cr2O7} in diluted \ch{H2SO4} or \ch{I2} in acetonitrile, a superconducting 2M-\ch{WS2} phase (Fig. S1 e,f), which consists of 1T'-\ch{WS2} layers with face-sharing distorted \ch{WS6} octahedra (Fig. S1 g,h) in a two-layer unit cell, can be synthesized (Fig. 1a)\cite{fang2019discovery,lai2021metastable}. This 2M-\ch{WS2} phase has the highest superconducting transition temperature ($T_c$) among the TMDs. In a monolayer, the potential combination of superconductivity and topology opens up a route to access non-Abelian states that are the key for topological quantum computing\cite{fu2008superconducting}. 

2M-\ch{WS2} can be mechanically exfoliated down to the monolayer limit in its structural 1T'-\ch{WS2} unit. The mechanically exfoliated 1T'-\ch{WS2} monolayer has been reported to be metallic; its resistivity drops at 5.7 K, but does not reach zero\cite{lai2021metastable}. Thus, it is unclear whether the monolayer is superconducting. 
 
Chemical exfoliation offers another route toward monolayers, with the advantage that it accesses large quantities of such, which can then be processed to printable inks, moving the studies from the lab setting to potential industrial application, especially if the synthesized ink is stable in air. It is well established that metallic \ch{WS2} monolayer nanosheets can be synthesized via Li intercalation of 2H-\ch{WS2} and subsequent sonication in water, but these nanosheets are never to 100 $\%$ in the 1T' phase, and usually have many defects\cite{yang1996li,tsai1997exfoliated,heising1999exfoliated,voiry2013enhanced,pierucci2019evidence,voiry2015covalent}. While such metallic \ch{WS2} has been studied extensively for catalytic applications\cite{voiry2013enhanced,cheng2014ultrathin}, to the best of our knowledge, it has never been investigated whether it is superconducting. In general, high-quality 2D superconducting monolayer suspensions are scarce. Chemical-exfoliated restacked-\ch{TaS2} nanosheets are superconducting with a a $T_c$ of 3 K\cite{pan2017enhanced}. 
1T'-\ch{MoS2} nanosheets show a $T_c$ of 4.6 K\cite{guo2017observation}, and recently reported printed, electrochemically exfoliated \ch{NbSe2} nanosheet films have a $T_c$ of 6.8 K\cite{li2021printable}. Out of these, only the last material has been shown to be usable as a printable ink. In this case, however, protective organic molecules are necessary to stabilize the ink, as \ch{NbSe2} is relatively air-sensitive\cite{wang2017high}. Furthermore, the ink was synthesized electrochemically, a method that is limited to metals\cite{li2021printable}, which would exclude 2H-\ch{WS2} as a starting material.

In this study, we report a simple chemical exfoliation method to make a stable superconducting ink containing 1T'-\ch{WS2} monolayers from \ch{K_{0.5}WS2}. We show that the sheets are stable in water, which provides a cheap, non-toxic and abundant ink-solvent for potential printable superconducting electronics. Exfoliation with high yield is then achieved by sonication, resulting in a suspension composed of monolayers with lateral sizes up to tens of micro-meters, which crystallize in the 1T'-structure. The composition and structure of the products are characterized with multiple diffraction, microscopy and spectroscopy techniques, establishing that the structure remains intact and low in defects, suggesting they are of much higher quality than their mechanical or Li-intercalation exfoliated counterparts. We prove that a thin film cast from the nanosheet ink is superconducting below 7.3 K, with an in-plane upper critical magnetic field of 30.1 T and an out-of-plane upper critical magnetic field of 5.3 T. The film shows highly anisotrpoic superconducting properties, that resemble these observed in gated 1T'-\ch{WTe2} pointing to 2D superconductivity and a potential exotic origin\cite{fatemi2018electrically,sajadi2018gate}. After exposing the printed film to ambient conditions for 30 days, its electronic transport behavior, as well as its Raman and X-ray photoelectron spectroscopy (XPS) spectra, remain unchanged. Finally, we show that, besides water, the exfoliated 1T'-\ch{WS2} monolayers can be well dispersed in several common solvents such as ethanol, iso-propanol (IPA), and dimethylformamide (DMF). The ink forms room-temperature conducting films on various known substrates, such as \ch{SiO2}/Si wafers, borosilicate glass, and indium tin oxide (ITO) coated glass, as well as flexible substrates such as polyethylene terephthalate (PET), polyethylene naphthalate (PEN), and silicone elastomer. Thus, the 1T'-\ch{WS2} monolayer-ink that we present here has a large application range, such as 3D printing, integrated circuits, and flexible devices.

\section*{Chemical exfoliation of 1T'-\ch{WS2} monolayers}
The starting compound \ch{K_{0.5}WS2} was synthesized via a solid state reaction, with full experimental details reported in the supplemental information. To the best of our knowledge, its crystal structure had not yet been reported. We resolved its structure by single crystal x-ray diffraction (SCXRD, Table S1), as shown in Fig. 1a. \ch{K_{0.5}WS2} crystallizes in the monoclinic space group \textit{C}2/\textit{m} and consists of layers of distorted \ch{WS6} octahedra that are structurally similar to the layers in 1T' (or $T_d$)-\ch{WTe2}\cite{brixner1962preparation,mar1992metal}. The K atoms in the interlayer space are disordered, and the EDS analysis shows that there is 0.5 K per formula unit in the crystals (Fig. S2).

We designed a route to chemically exfoliate \ch{K_{0.5}WS2} to 1T'-\ch{WS2} monolayers directly in acid (Fig. 1a). Details of the process can be found in the SI. A highly stable nanosheet ink in MilliQ water can be obtained if large unexfoliated pieces are removed via centrifugation at 2000 rpm as indicated by a zeta potential of -57.5$\pm$4 mV\cite{kumar2017methods}. The nanosheets presented here are stable in water without organic surfactant molecules whose presence is known to hinder the application of the as-synthesized nanomaterials for electronic purposes\cite{lin2019van}. 

\section*{Structural characterization of 1T'-\ch{WS2} monolayers}
The diluted nanosheet suspension was deposited on a silicon wafer and the sheets were characterized with atomic force microscopy (AFM). As shown in Fig. 1b, the exfoliated 1T'-\ch{WS2} nanosheets have a thickness of about 0.7 nm if measured on top of another nanosheet, which agrees well with the monolayer thickness of 1T'-\ch{WS2}. The nanosheet is 1.2 nm thin if measured on the wafer directly, which is a common value for chemically exfoliated TMD monolayers on wafers, due to absorbed water molecules\cite{lerf1977solvation,weber2018irooh,song2019soft,ferrenti2019change}. Fig. 1c shows a typical TEM image of 1T'-\ch{WS2} monolayers randomly stacked on top of each other. The selected area electron diffraction (SAED) on a free-standing monolayer is shown in Fig. 1d, confirming its high crystallinity and the 1T' structure (Fig. S3). An atomic resolution scanning transmission electron microscopy (STEM) image of a monolayer 1T'-\ch{WS2} is shown in Fig. 1e, with no visible defects and impurity phases. Fig. 1f shows the zigzag chains of W atoms of a typical 1T'-TMD structure (Fig. 1g,h). S atoms cannot be resolved in Fig. 1f, as the STEM imaging is a Z-contrast technique and S has a much smaller atomic number compared with W. We confirm the existence of S and its relative ratio to W in the monolayers by EDS (Fig. S4). Our AFM and TEM analysis found that the ink seems to be almost entirely composed of monolayers and that all larger unexfoliated pieces could be successfully removed with centrifugation as mentioned above. Finally, in order to differentiate these monolayers form their semiconducting 1H counterparts, we performed EELS studies on both, a monolayer 1T'-\ch{WS2} and a monolayer 1H-\ch{WS2}. The valence EELS spectra are shown in Fig. 1i. In the case of 1H-\ch{WS2}, features appear around 2, 2.4, and 3 eV, which are the A, B, and C excitons that are associated to the electronic properties of the semiconducting phase\cite{van2021illuminating,molina2013effect,qiu2013optical,nerl2017probing}, and the broad peak around 8 eV, which corresponds to the $\pi$ plasmon\cite{nerl2017probing,moynihan2020plasmons,marinopoulos2004ab}. In contrast, the single-layer 1T'-\ch{WS2}, does not exhibit exciton features observed in the 1H-phase; instead, only one broad peak around 6 eV can likely be attributed to a $\pi$ plasmon. This clearly distinguishes the electronic properties of 1T'-\ch{WS2} from its semiconducting 1H counterpart.

\subsection*{Structure and unconventional superconductivity of the printed 1T'-\ch{WS2} film}

\begin{figure}[H]
\centering
  \includegraphics[width=1.0\textwidth]{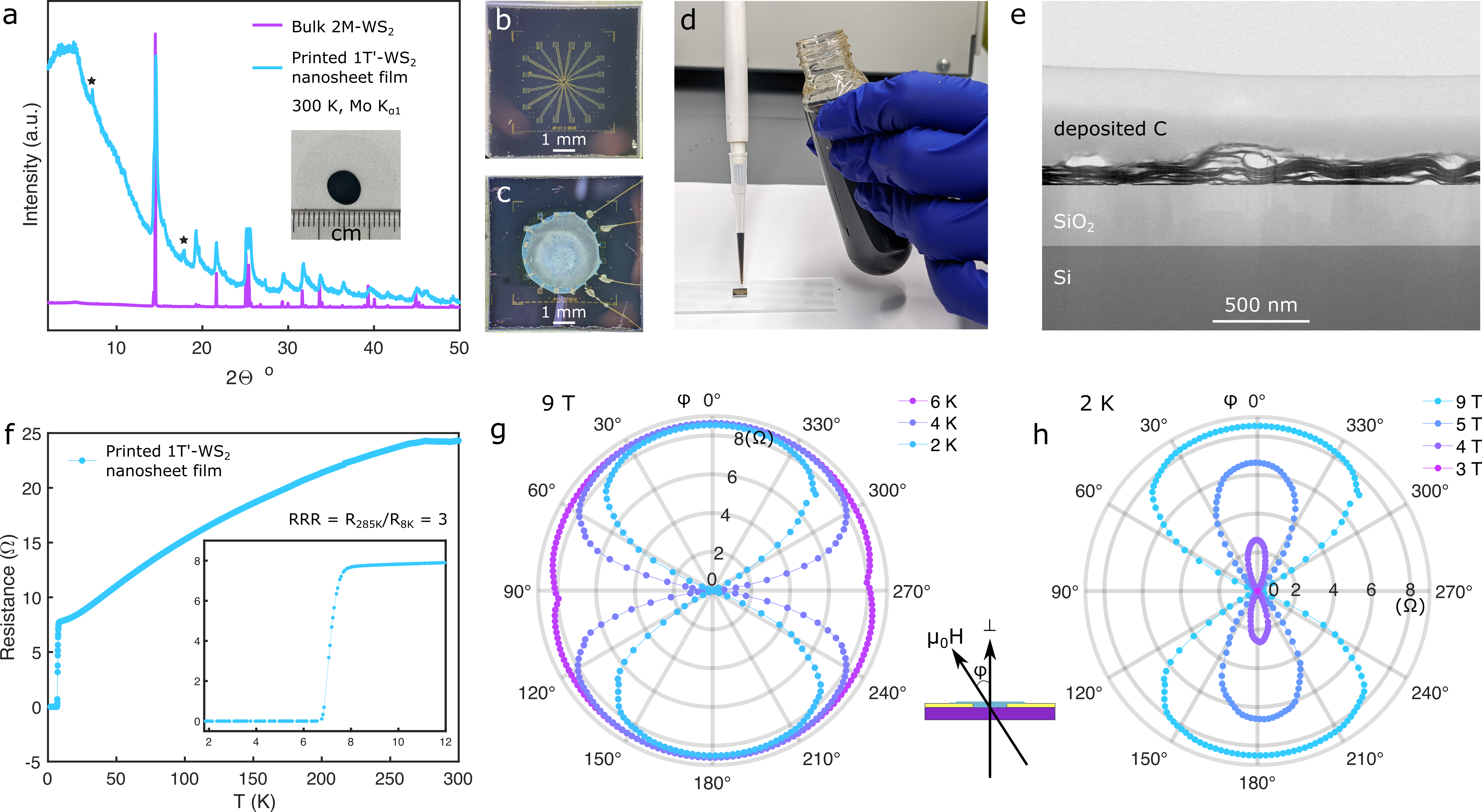}
  \caption{(a) The 1T'-\ch{WS2} nanosheet-ink was deposited on a polymer film and dried (inset) for an in-plane powder XRD characterization. The nanosheet PXRD pattern is compared to the in-plane pattern of bulk 2M-\ch{WS2}. The stars indicate additional peaks due to sheets restacking and the contribution from out-of-plane diffraction. (b) A \ch{SiO2}/Si wafer with preprinted electrodes; (c) 1T'-\ch{WS2} nanosheet film deposited on the \ch{SiO2}/Si wafer shown in (b). (d) A pipette can be used to deposit the nanosheet-ink. (e) A bright field STEM image showing the cross-sectional structure of the printed 1T'-\ch{WS2} nanosheet film on a \ch{SiO2}/Si wafer. (f) Resistance versus temperature data from 300 K to 1.8 K for a freshly printed 1T'-\ch{WS2} film without any external magnetic field. The inset shows the superconducting transition region. (g) Angle-dependent resistance data of the printed 1T'-\ch{WS2} device measured from 2 K to 6 K with a 9 T external magnetic field when the field is rotated from perpendicular to parallel to the device plane. (h) Angle-dependence of the resistance measured at 2 K with a 3 T to 9 T external magnetic field (Inset: An illustration showing the experimental configuration where the magnetic field is perpendicular to the printed device plane at $\varphi$ = 0 $^\circ$ and $\varphi$ = 360 $^\circ$).}
  \label{figure2}
  \centering
\end{figure}

Having established that we can produce an ink made of metallic 1T'-\ch{WS2} monolayers, we can now study its properties. The ink was first deposited and dried on a polymer film, as shown in the inset of Fig. 2a. The structure of the dried film was characterized by in-plane PXRD in transmission mode. A 2M-\ch{WS2} crystal was measured in the same way for comparison. As shown in Fig. 2a, the patterns align, suggesting the sheets retained good crystallinity. Two broad peaks appear in addition in the pattern of the ink (labeled with stars), these come from some out-of-plane contribution of crumbled sheets in the printed film\cite{li2005positively,omomo2003redoxable,sasaki1998osmotic}. The Raman spectrum of the printed 1T'-\ch{WS2} films have the characteristic peaks of the bulk 2M-\ch{WS2} with extra peaks showing up at 196 cm$^{-1}$ and 400 cm$^{-1}$, which can be attributed to the loss of symmetry in the monolayers (Fig. S5).

To study the electronic transport properties of the 1T'-\ch{WS2} nanosheet-ink, a droplet was deposited on a silicon wafer with pre-patterned electrodes as shown in Fig. 2b--d. The ink droplet was dried in ambient conditions before Au wires were attached to the exposed pre-patterned electrodes as shown in Fig. 2c. 
To gain an insight of how the nanosheets deposit on the wafer, we cut a sample of a dried nanosheet film on a silicon wafer with a focused ion beam (FIB) and studied its cross section with STEM. A typical bright field image is shown in Fig. 2e. The film shows that the sheets are in good contact and that the nanosheets randomly stack on top of each other. Even though in some areas, the sheets are crumbled, the majority of the sheets are well-oriented. 

The temperature ($T$)-dependent resistance ($R$) in Fig. 2f shows that the device is metallic, as the resistance decreases with decreasing temperature. At $\sim$7.7 K, the resistance drops sharply and reaches zero at $\sim$6.6 K (Fig. 2f: inset). We define the $T_c$ as the temperature where the resistance drops to 50 \% of the normal state resistance, which is 7.3 K. As the film is 2D in nature, a strong anisotropy with respect to an applied magnetic field ($\mu_{0}H$) direction can be expected. The angle-dependent resistance under an applied magnetic field of 9 T at different temperatures is shown in Fig. 2g. Similarly, the angle-dependent resistance at 2 K with different magnetic field strengths is shown in Fig. 2h. The external magnetic field is applied perpendicular to the device plane at $\varphi$ = 0$^\circ$ and 180$^\circ$ (out-of-plane, $\mu_{0}H^\perp$), and it is parallel to the device plane at $\varphi$ = 90$^\circ$ and 270$^\circ$ (in-plane, $\mu_{0}H^{||}$) (inset between Fig. 2g and 2h). The electronic transport is highly anisotropic in the superconducting state; it is more easily suppressed when $\mu_{0}H^\perp$, and more robust when $\mu_{0}H^{||}$ (Fig. 2g,h). The angle-dependent resistance data shown in Fig. 2g,h suggest that the resistance signal stems predominately from well-orientated nanosheets despite of some crumbling seen in the cross-sectional image (Fig. 2e).

\begin{figure}[H]
\centering
  \includegraphics[width=1.0\textwidth]{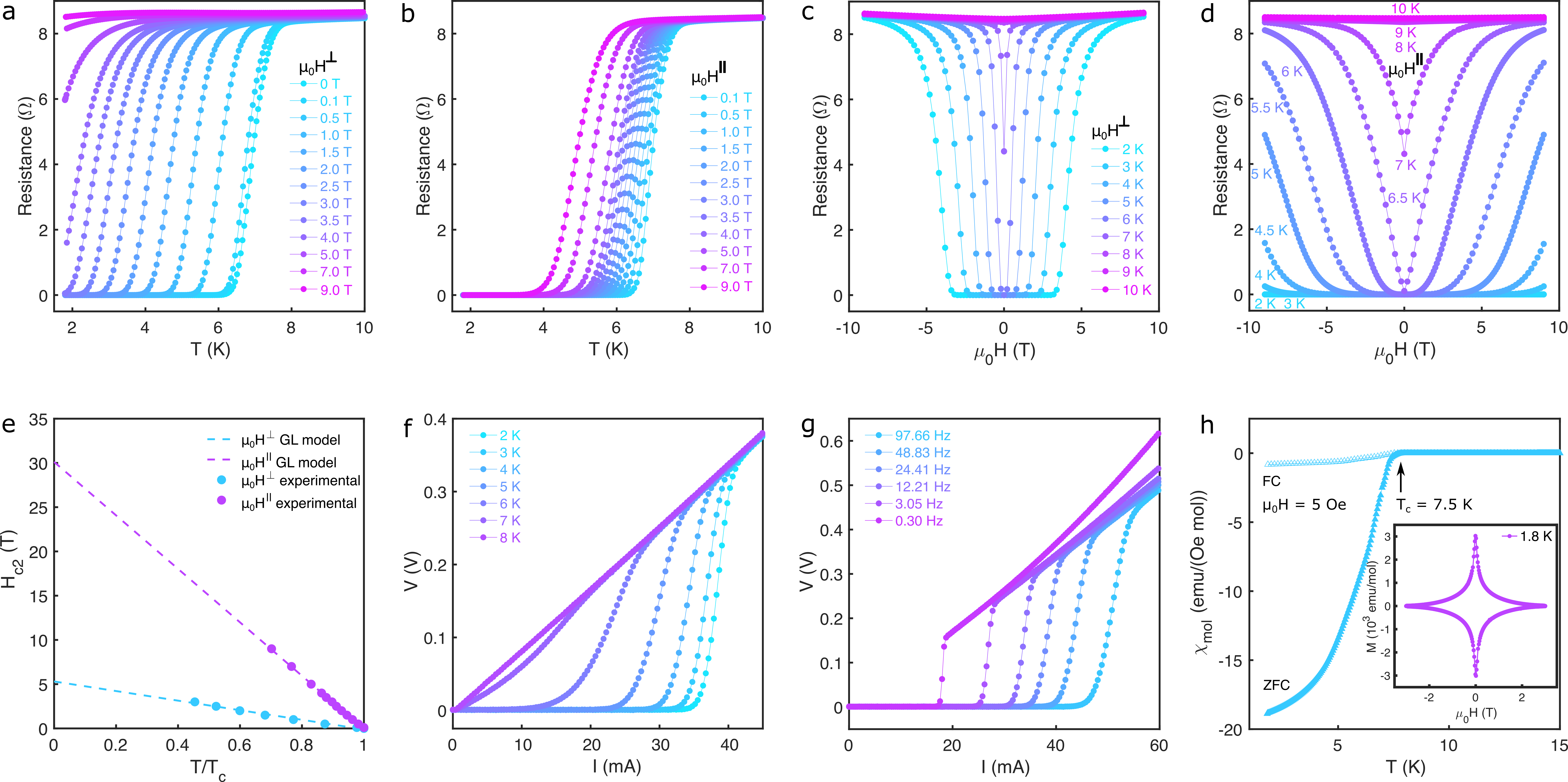}
  \caption{(a), (b) Temperature-dependent resistance ($R$-$T$) of the printed 1T'-\ch{WS2} film, measured under external magnetic fields ranging from 0 T to 9 T, applied perpendicular (a) or parallel (b) to the device plane. (c), (d) Isotherms of the printed 1T'-\ch{WS2} film from 2 K to 10 K, measured with an external magnetic field that is applied perpendicular (c) or parallel (d) to the device plane. (e) Upper critical field $H_{c2}$ versus $T_c$ plot for both $\mu_{0}H^\perp$ and $\mu_{0}H^{||}$. The experimental data are fitted using Ginzburg-Landau (GL) theory. (f) Current ($I$) vs. voltage ($V$) curves of the printed 1T'-\ch{WS2} film measured from 2 K to 8 K with an ac current of frequency 24.41 Hz. (g) $I$-$V$ curves measured at 2 K with different ac current frequencies. (h) Temperature-dependent magnetic susceptibility of dried 1T'-\ch{WS2} nanosheet-powder collected from the ink. (Inset: magnetic field-dependent magnetization of the same sample measured at 1.8 K.)}
  \label{figure3}
  \centering
\end{figure}

$R$-$T$ curves at different applied magnetic fields, both with $\mu_{0}H^\perp$ and $\mu_{0}H^{||}$, are shown in Fig. 3a,b. The resistance still drops to zero at the highest magnetic field of 9 T if it is applied along the in-plane direction.
Fig. 3c,d show the out-of-plane ($R$-$\mu_{0}H^\perp$ isotherms) and in-plane ($R$-$\mu_{0}H^{||}$ isotherms) field-dependent resistance of the printed 1T'-\ch{WS2} film around the transition temperature. Below $T_c$, the $R$-$\mu_{0}H^\perp$ isotherms (Fig. 3c) show a broad transition from the superconducting state to the normal state, and their corresponding critical magnetic field decreases as the temperature increases. When $\mu_{0}H$ is applied parallel to the device plane, below 3 K, the resistance remains zero when the applied magnetic field increases to 9 T, suggesting a very high critical magnetic field when applied parallel to the film. 
To determine the upper critical magnetic field ($H_{c2}$), the transition temperature at each applied magnetic field, corresponding to half of its normal state resistance, is plotted vs. the field, for both $\mu_{0}H^\perp$ and $\mu_{0}H^{||}$ (Fig. 3e). A linear correlation of $H_{c2}$ vs. $T_c$ can be modelled by the 2D Ginzburg-Landau (GL) theory\cite{lai2021metastable,li2021printable} for both directions:

\begin{equation*}
    H_{c2}(T) = \frac{\Phi_0}{2\pi\xi^2_{GL}(0)}(1-\frac{T}{T_c})
\end{equation*} 

where $\Phi_0$ is the magnetic flux quantum, and the $\xi_{GL}(0)$ is the zero-temperature GL in-plane coherence length. This results in an out-of-plane upper critical magnetic field ($H_{c2}^\perp(0)$) of 5.3 T and an in-plane GL superconducting coherence length of $\xi_{GL}(0)$ $\sim$7.9 nm. Fitting the in-plane $H_{c2}$ vs. $T_c$ yields an in-plane upper critical magnetic field ($H_{c2}^{||}(0)$) of 30.1 T. Similar to the recently reported printed \ch{NbSe2} film\cite{li2021printable}, the $H_{c2}^{||}(0)$ of the printed 1T'-\ch{WS2} film is very high, far beyond its BCS Pauli paramagnetic limit of 13.1 T ($H_p$ $\sim$1.84 $T_c$)\cite{clogston1962upper}. 
However, the symmetry of the centrosymmetric 1T' structure of \ch{WS2} is fundamentally different from the non-centrosymmetric hexagonal structure of \ch{NbSe2} where Ising type superconductivity is responsible for exceeding the Pauli limit\cite{li2021printable,xi2016ising}. A similar anisotropy of the critical field, where the in-plane critical field exceeds the Pauli limit, has been observed in 1T'-WTe$_2$. It was pointed out that the exact mechanism for this behaviour in the 1T' crystal structure is not understood\cite{fatemi2018electrically,sajadi2018gate}. An unconventional reason seems to be possible\cite{cao2018unconventional}, which should be probed in future studies on the monolayers.

The current ($I$) vs. voltage ($V$) curves of the device, measured with a fixed alternative current (ac) frequency (24.41 Hz), are shown in Fig. 3f for different temperatures. A critical current ($I_{c}$) $\sim$33 mA can be extracted at 2 K. The critical current decreases with increasing temperature, and the supercurrent eventually disappears at temperatures above $T_c$. When the frequency is varied, as shown in Fig. 3g, the critical current of the printed 1T'-\ch{WS2} film also changes. At 2 K, $I_{c}$ reaches a maximum of $\sim$44 mA with an excitation current frequency of 97.66 Hz. On the other hand, it decreases to $\sim$17 mA with the lowest excitation current frequency of 0.30 Hz. The $I$-$V$ curves become non-linear above $I_{c}$ for all the frequencies as the Joule heating appears.

A second device (device 2) was printed with a different batch of 1T'-\ch{WS2} nanosheet-ink to confirm the robustness of the cast superconducting films. The electronic transport data of device 2 are nearly identical and details are shown in the SI (Fig. S6).

Next, we studied the dc magnetic susceptibility of a re-stacked nanosheet pellet that was collected from the dried ink (Fig. 3h). A strong diamagnetic signal is observed below $T_c$ = 7.5 K under zero field cooled (ZFC) condition. Field cooling (FC) with an applied magnetic field of 5 Oe suppresses the diamagnetic response due to the Meissner-Ochsenfeld effect. The $\chi_{mol}$ of the nanosheet pellet at 2 K is within the same order of magnitude than bulk 2M-\ch{WS2} (Fig. S7), thus the majority volume of the restacked sample is superconducting. Both the electronic as well as the magnetic characterizations show that the ink is composed of high-quality superconducting 1T'-\ch{WS2} nanosheets that are ready to be used for printing electronics on various substrates.

\begin{figure}[H]
\centering
  \includegraphics[width=0.6\textwidth]{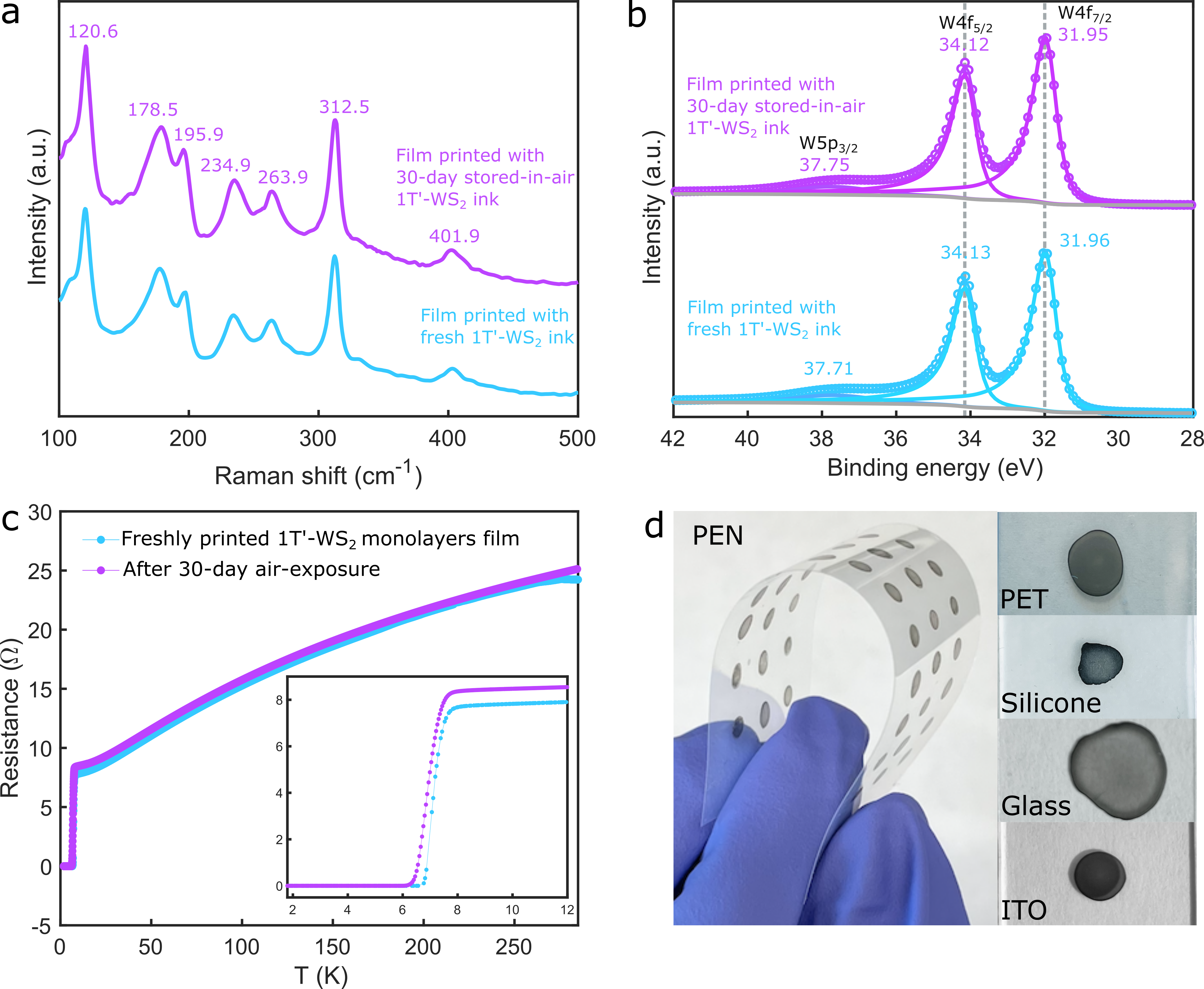}
  \caption{Raman spectra (a), W XPS spectra (b) of the films printed with newly synthesized 1T'-\ch{WS2} nanosheet-ink and the 1T'-\ch{WS2} nanosheet-ink that is stored in air for a month. (c) $R$-$T$ curves of the printed 1T'-\ch{WS2} film measured right after preparing the device and after storing in air for a month. (d) Patterns printed with the 1T'-\ch{WS2} nanosheet-ink on a PEN film, PET, silicone, glass, and ITO. No obvious cracks or fallen off pieces of the printed patterns is observed while folding the PEN substrate (left), suggesting a good affinity of the 1T'-\ch{WS2} nanosheet to the flexible substrate.}
  \label{figure4}
  \centering
\end{figure}

\section*{Air-stability}
To investigate the air-stability of the 1T'-\ch{WS2} nanosheet-ink and the printed device, we stored the ink as well as both devices in ambient conditions for a month. A new nanosheet film was printed from the air-exposed ink, and its Raman spectrum is identical to the spectrum from the freshly printed film (Fig. 4a). The W and S XPS spectra of both films are also identical, suggesting that no oxidation or phase transformation appears when the 1T'-\ch{WS2} nanosheet-ink is stored in ambient atmosphere (Fig. 4b, Fig. S8). The W Spectrum is fitted with one set of W4f doublets with W4f$_{7/2}$ at 31.95 eV and W4f$_{5/2}$ at 34.12 eV, proving that the nanosheets are purely in the 1T' phase. Temperature-dependent resistance measurements on device 1 after one month of air exposure show that the device has almost the same room temperature resistance, the same residual resistivity ratio (RRR=$R_{285 K}$/$R_{8 K}$) $\sim$3, and the $T_c$ is almost unchanged as compared to the fresh sample (Fig. 4c). This is different compared to most other known 2D materials that require preparation in an inert environment and the protection of organic molecules to be handled in air\cite{li2021printable,cao2015quality,gray2020cleanroom,yasaei2015high,kang2015solvent}. The 1T'-\ch{WS2} nanosheet-ink presented here is robust in water, and is stable in ambient conditions without protection, which gives this ink a higher potential for real-world applications.

\section*{Dispersity in different solvents and printability on various substrates}
Finally, we tested if an ink can be created with solvents other than water as well as the variety of substrates the ink can be cast on. Common solvents such as hexane, methanol, ethanol, IPA, acetone, acetonitrile, DMF, THF, and DMSO were used to disperse the sheets (Fig. S9). The nanosheets can be dispersed in ethanol, IPA, and DMF. Among all tested solvents, water gives the best dispersity and highest stability of the 1T'-\ch{WS2} nanosheets. The aqueous ink was deposited on various substrates, including hard substrates such as \ch{SiO2}/Si wafers, borosilicate glass, and ITO coated glass, as well as flexible substrates such as PET, PEN, and the silicone elastomer (Fig. 4d). The room-temperature conductivity of the printed nanosheets patterns on the different substrates was measured with a multimeter, showing that each film is metallic at room temperature. Thus, the 1T'-\ch{WS2} monolayer-ink can be prepared with various solvents and can be printed on many different substrates, expanding its possible applications to integrated circuits, and flexible devices.

\section*{Conclusion}
In conclusion, we successfully synthesized monolayers of the 2D TI candidate 1T'-\ch{WS2} and prepared an air-stable aqueous superconducting ink consisting primarily of 1T'-\ch{WS2} monolayers. A printed 1T'-\ch{WS2} film is metallic at room temperature, and superconducting below 7.3 K with a maximal critical current of 44 mA at 2 K. The upper critical magnetic field of 30.1 T if the field is applied in plane, and 5.3 T for the perpendicular field direction, pointing to unconventional superconductivity. Both the 1T'-\ch{WS2} monolayer-ink, as well as the printed film, are stable at ambient conditions for at least 30 days, without any protective agents. This also suggests that a monolayer 1T'-\ch{WS2} is air stable and superconducting, opening up avenues for investigating the interplay between topology and superconductivity in this 2D material. We further show that the ink can form conducting films on various substrates, such as \ch{SiO2}/Si wafer, borosilicate glass, ITO, PET, PEN, and the silicone elastomer. The simple synthesis, stability and versatility of the ink reported here suggest that it might find applications in several areas, such as quantum computing, integrated circuit to flexible and wearable devices. Furthermore the air-stable monlayers can be studied for their potential interplay of topology and superconductivity.

\bibliographystyle{Science}

\begin{thebibliography}{10}

\bibitem{torrisi2014electrifying}
F.~Torrisi, J.~N. Coleman, {\it Nature nanotechnology\/} {\bf 9}, 738 (2014).

\bibitem{lin2018solution}
Z.~Lin, {\it et~al.\/}, {\it Nature\/} {\bf 562}, 254 (2018).

\bibitem{eda2008large}
G.~Eda, G.~Fanchini, M.~Chhowalla, {\it Nature nanotechnology\/} {\bf 3}, 270
  (2008).

\bibitem{shi2017highly}
P.~Shi, {\it et~al.\/}, {\it Advanced Materials\/} {\bf 29}, 1703455 (2017).

\bibitem{kelly2021electrical}
A.~G. Kelly, D.~O'Suilleabhain, C.~Gabbett, J.~N. Coleman, {\it Nature Reviews
  Materials\/} pp. 1--18 (2021).

\bibitem{li2021printable}
J.~Li, {\it et~al.\/}, {\it Nature Materials\/} {\bf 20}, 181 (2021).

\bibitem{wu2018observation}
S.~Wu, {\it et~al.\/}, {\it Science\/} {\bf 359}, 76 (2018).

\bibitem{fatemi2018electrically}
V.~Fatemi, {\it et~al.\/}, {\it Science\/} {\bf 362}, 926 (2018).

\bibitem{sajadi2018gate}
E.~Sajadi, {\it et~al.\/}, {\it Science\/} {\bf 362}, 922 (2018).

\bibitem{qian2014quantum}
X.~Qian, J.~Liu, L.~Fu, J.~Li, {\it Science\/} {\bf 346}, 1344 (2014).

\bibitem{coleman2011two}
J.~N. Coleman, {\it et~al.\/}, {\it Science\/} {\bf 331}, 568 (2011).

\bibitem{fang2019discovery}
Y.~Fang, {\it et~al.\/}, {\it Advanced Materials\/} {\bf 31}, 1901942 (2019).

\bibitem{lai2021metastable}
Z.~Lai, {\it et~al.\/}, {\it Nature Materials\/} pp. 1--8 (2021).

\bibitem{fu2008superconducting}
L.~Fu, C.~L. Kane, {\it Physical review letters\/} {\bf 100}, 096407 (2008).

\bibitem{yang1996li}
D.~Yang, R.~Frindt, {\it Journal of Physics and Chemistry of Solids\/} {\bf
  57}, 1113 (1996).

\bibitem{tsai1997exfoliated}
H.-L. Tsai, J.~Heising, J.~L. Schindler, C.~R. Kannewurf, M.~G. Kanatzidis,
  {\it Chemistry of materials\/} {\bf 9}, 879 (1997).

\bibitem{heising1999exfoliated}
J.~Heising, M.~G. Kanatzidis, {\it Journal of the American Chemical Society\/}
  {\bf 121}, 11720 (1999).

\bibitem{voiry2013enhanced}
D.~Voiry, {\it et~al.\/}, {\it Nature materials\/} {\bf 12}, 850 (2013).

\bibitem{pierucci2019evidence}
D.~Pierucci, {\it et~al.\/}, {\it Applied Physics Letters\/} {\bf 115}, 032102
  (2019).

\bibitem{voiry2015covalent}
D.~Voiry, {\it et~al.\/}, {\it Nature chemistry\/} {\bf 7}, 45 (2015).

\bibitem{cheng2014ultrathin}
L.~Cheng, {\it et~al.\/}, {\it Angewandte Chemie\/} {\bf 126}, 7994 (2014).

\bibitem{pan2017enhanced}
J.~Pan, {\it et~al.\/}, {\it Journal of the American Chemical Society\/} {\bf
  139}, 4623 (2017).

\bibitem{guo2017observation}
C.~Guo, {\it et~al.\/}, {\it Journal of Materials Chemistry C\/} {\bf 5}, 10855
  (2017).

\bibitem{wang2017high}
H.~Wang, {\it et~al.\/}, {\it Nature communications\/} {\bf 8}, 1 (2017).

\bibitem{brixner1962preparation}
L.~H. Brixner, {\it Journal of Inorganic and Nuclear Chemistry\/} {\bf 24}, 257
  (1962).

\bibitem{mar1992metal}
A.~Mar, S.~Jobic, J.~A. Ibers, {\it Journal of the American Chemical Society\/}
  {\bf 114}, 8963 (1992).

\bibitem{kumar2017methods}
A.~Kumar, C.~K. Dixit, {\it Advances in nanomedicine for the delivery of
  therapeutic nucleic acids\/} (Elsevier, 2017), pp. 43--58.

\bibitem{lin2019van}
Z.~Lin, Y.~Huang, X.~Duan, {\it Nature Electronics\/} {\bf 2}, 378 (2019).

\bibitem{lerf1977solvation}
A.~Lerf, R.~Sch{\"o}llhorn, {\it Inorganic Chemistry\/} {\bf 16}, 2950 (1977).

\bibitem{weber2018irooh}
D.~Weber, {\it et~al.\/}, {\it Journal of Materials Chemistry A\/} {\bf 6},
  21558 (2018).

\bibitem{song2019soft}
X.~Song, {\it et~al.\/}, {\it Journal of the American Chemical Society\/} {\bf
  141}, 15634 (2019).

\bibitem{ferrenti2019change}
A.~M. Ferrenti, {\it et~al.\/}, {\it Inorganic Chemistry\/} {\bf 59}, 1176
  (2019).

\bibitem{van2021illuminating}
S.~E. van Heijst, {\it et~al.\/}, {\it Annalen der Physik\/} {\bf 533}, 2000499
  (2021).

\bibitem{molina2013effect}
A.~Molina-S{\'a}nchez, D.~Sangalli, K.~Hummer, A.~Marini, L.~Wirtz, {\it
  Physical Review B\/} {\bf 88}, 045412 (2013).

\bibitem{qiu2013optical}
D.~Y. Qiu, H.~Felipe, S.~G. Louie, {\it Physical review letters\/} {\bf 111},
  216805 (2013).

\bibitem{nerl2017probing}
H.~C. Nerl, {\it et~al.\/}, {\it npj 2D Materials and Applications\/} {\bf 1},
  1 (2017).

\bibitem{moynihan2020plasmons}
E.~Moynihan, {\it et~al.\/}, {\it Journal of Microscopy\/} {\bf 279}, 256
  (2020).

\bibitem{marinopoulos2004ab}
A.~Marinopoulos, L.~Reining, A.~Rubio, V.~Olevano, {\it Physical Review B\/}
  {\bf 69}, 245419 (2004).

\bibitem{li2005positively}
L.~Li, R.~Ma, Y.~Ebina, N.~Iyi, T.~Sasaki, {\it Chemistry of materials\/} {\bf
  17}, 4386 (2005).

\bibitem{omomo2003redoxable}
Y.~Omomo, T.~Sasaki, L.~Wang, M.~Watanabe, {\it Journal of the American
  Chemical Society\/} {\bf 125}, 3568 (2003).

\bibitem{sasaki1998osmotic}
T.~Sasaki, M.~Watanabe, {\it Journal of the American Chemical Society\/} {\bf
  120}, 4682 (1998).

\bibitem{clogston1962upper}
A.~M. Clogston, {\it Physical Review Letters\/} {\bf 9}, 266 (1962).

\bibitem{xi2016ising}
X.~Xi, {\it et~al.\/}, {\it Nature Physics\/} {\bf 12}, 139 (2016).

\bibitem{cao2018unconventional}
Y.~Cao, {\it et~al.\/}, {\it Nature\/} {\bf 556}, 43 (2018).

\bibitem{cao2015quality}
Y.~Cao, {\it et~al.\/}, {\it Nano letters\/} {\bf 15}, 4914 (2015).

\bibitem{gray2020cleanroom}
M.~J. Gray, {\it et~al.\/}, {\it Review of Scientific Instruments\/} {\bf 91},
  073909 (2020).

\bibitem{yasaei2015high}
P.~Yasaei, {\it et~al.\/}, {\it Advanced Materials\/} {\bf 27}, 1887 (2015).

\bibitem{kang2015solvent}
J.~Kang, {\it et~al.\/}, {\it ACS nano\/} {\bf 9}, 3596 (2015).

\end{thebibliography}

\section*{Acknowledgments}
\textbf{Funding:} This research was supported by DOD ONR (N00014-21-1-2733), NSF-MRSEC program (DMR-2011750), the Gordon and Betty Moore Foundation (GBMF9064). The authors acknowledge the use of Princeton's Imaging and Analysis Center, which is partially supported by the Princeton Center for Complex Materials, a National Science Foundation (NSF)-MRSEC program (DMR-2011750). \textbf{Acknowledgments:} The authors thank Professor Shengwei Jiang from Shanghai Jiao Tong University, Professor Jie Shan and Professor Kin Fai Mak from Cornell University for providing us the wafers with the prepatterned electrodes. The authors thank Fei Gao and Tianran Liu from the department of electrical and computer engineering of Princeton University for their helpful discussions about integrated circuits and flexible substrates. The authors thank all the members from the Schoop lab at Princeton University for their helpful discussions about this manuscript. \textbf{Pending patent application:} No. 63/327,103 titled “Superconducting \ch{WS2}-Based Nanosheet Ink for Printable and Flexible Electronics”. \textbf{Author Contributions:} X.S. and L.M.S. conceived the project. X.S. synthesized the materials, prepared the printed devices and conducted most of the materials characterization, such as SEM/EDS, AFM, TEM, DP, PXRD, Raman, XPS, zeta potential measurements. X.S. tested the dispersity of the nanosheets in different solvents and studied the film formation on various substrates. Transport and magnetization measurements were carried out by R.S and X.S.. G.C. and X.S. carried out the FIB sample preparation, HAADF-STEM and EDS characterization under the supervision of N.Y.. Y.Y. conducted the EELS measurement and analyzed the data under the supervision of P.E.B.. F.K. conducted the Raman-spectroscopy calculation under the supervision of F.P.. J.F.K. carried out the SCXRD of the parent compound \ch{K_{0.5}WS2} and resolved its structure. X.S. analyzed the experimental data under the supervision of L.M.S. X.S. and L.M.S. wrote the initial draft of the manuscript. All authors discussed the results and provided input to the manuscript. \textbf{Competing interests:} All authors listed on this manuscript declare no competing interests. \textbf{Data and materials availability:} All data is available in the manuscript or the supplementary materials.

\section*{Supplementary materials}
Materials and Methods\\
Figs. S1 to S9\\
Tables S1\\
References

\end{document}